# Simultaneously high electron and hole mobilities in cubic boron-V compounds: BP, BAs and BSb

Te-Huan Liu[1], Bai Song[1], Laureen Meroueh[1], Zhiwei Ding[1], Qichen Song[1], Jiawei Zhou[1], Mingda Li[2], and Gang Chen[1,*]

[1]*Department of Mechanical Engineering, Massachusetts Institute of Technology, Cambridge, MA, 02139, USA*

[2]*Department of Nuclear Science and Engineering, Massachusetts Institute of Technology, Cambridge, MA, 02139, USA*

## Abstract

Through first-principles calculations, the phonon-limited transport properties of cubic boron-V compounds (BP, BAs and BSb) are studied. We find that the high optical phonon frequency in these compounds leads to the substantial suppression of polar scattering and the reduction of inter-valley transition mediated by large-wavevector optical phonons, both of which significantly facilitate charge transport. We also discover that BAs simultaneously has a high hole mobility (2110 cm$^2$/V-s) and electron mobility (1400 cm$^2$/V-s) at room temperature, which is rare in semiconductors. Our findings present a new insight in searching high mobility polar semiconductors, and point to BAs as a promising material for electronic and photovoltaic devices in addition to its predicted high thermal conductivity.

---

[*] Author to whom correspondence should be addressed: gchen2@mit.edu

**Main text**

Carrier mobility is key to the performance of modern electronic devices such as diodes, transistors, photovoltaic cells, and thermoelectric modules [1]. After Bloch's [2] pioneering study of the interaction between electrons and lattice vibrations, it was soon realized that in common metals and semiconductors, electrons are predominantly scattered by phonons [3]. To describe the coupling between electrons and long-wavelength acoustic phonons, Bardeen and Shockley [4] introduced the concept of deformation potential, through which the energy offset at the electronic band edge is related to the lattice strain caused by in-phase atomic displacement. In non-polar semiconductors, the acoustic-deformation-potential scattering often dominates the electron-phonon interaction. Fröhlich [5] illustrated that the macroscopic polarization field generated by the out-of-phase vibration (longitudinal-optical (LO) phonon) in electric dipoles tends to strongly couple with electrons. This is called the polar-optical-phonon scattering, and acts as one significant scattering channel to electrons in polar crystals. In addition to these two types of electron-phonon interactions, there are two more scattering mechanisms: optical-deformation-potential scattering, an analogy of acoustic-deformation-potential scattering to transverse-optical (TO) phonons first investigated by Herring *et al.* [6]; and piezoelectric scattering, an analogy of polar-optical-phonon scattering to longitudinal-acoustic (LA) phonons first investigated by Meijer *et al.* [7].

Based on the aforementioned phenomenological pictures, electron-phonon coupling matrices can be derived as explicit, semi-empirical formulae, and the electron scattering rates can thus be evaluated using Fermi's golden rule [8]. Such calculations have been performed by Rode *et al.* in calculating the electron mobility of IV [9], III-V [10,11] and II-VI [12] group semiconductors, by solving the Boltzmann transport equation (BTE). In addition, associated with use of Monte Carlo technique, Fischetti and Laux investigate the electron mobilities in different types of materials such as Si, GaAs and SiGe alloys [13-15].

The recent development of density-functional perturbation theory (DFPT) [16] has opened a new route to investigating electron-phonon interactions within the first-principles framework. In addition, the Wannier interpolation scheme [17,18] has enabled electron transport studies through a mode-by-mode analysis on a very fine **k**-mesh, based on first-principles inputs. As a result of these advancements, computation of transport properties by combining first-principles coupling matrices with BTE becomes practical when momentum space can be near continuously sampled. Carrier transport has thus been studied in a series of benchmark materials. For silicon, Restrepo *et al.* [19] initially carried out the first-principles calculation for the electron mobility. Qiu *et al.* [20] then presented carrier mobility and mean-free-path (MFP) spectra at different carrier concentrations and temperatures. The electron-phonon interactions as well as the transport properties of two-dimensional materials such as graphene [21], silicene [22], phosphorene [23], and transition metal dichalcogenides [24] were also presented. Very recently, this framework was extended to study carrier mobility in the strongly polar material GaAs [25,26], and to study the thermoelectric properties of IV-VI group [27,28] and half-Heusler compounds [29].

Although one of the least studied III-V semiconductors, zinc-blende BAs has recently drawn much attention due to its predicted high intrinsic thermal conductivity [30]. Theoretical studies have revealed that the room-temperature thermal conductivity of BAs could reach as high as ~1400 W/m-K [31]. Meanwhile, experimentally obtained thermal conductivities in the range of only 200 to 400 W/m-K have been reported [32-35], primarily due to challenges in growing large defect-free BAs crystals. Nonetheless, it is already a promising competitor to the best traditional heat conductors such as copper, and further improvement in crystal quality is in progress. Based on the Shockley-Queisser theory [36,37], BAs has also been considered as a candidate for photovoltaic material owing to a properly 1.50 eV bandgap [38,39]. In addition, it was recently suggested via first-principles calculations that BAs has much longer hot-carrier lifetime than other IV and III-V group semiconductors [40], which further benefits its use in photovoltaics.

Despite the previous efforts devoted in studying transport properties, little is known about the carrier mobilities as well as the carrier MFPs in single-crystal BAs. The highest experimentally observed hole mobility of BAs is about 400 cm$^2$/V-s [41,42], which may have been limited by the quality of available samples. In this work, we present first-principles calculations of charge transport properties of single-crystal zinc-blende boron compounds including BAs, BP and BSb at different temperatures and carrier concentrations. We start with the scattering rate of electrons upon electron-phonon interaction, which can be derived via Fermi's golden rule under the relaxation time approximation [8]

$$\frac{1}{\tau_{n\mathbf{k}}^{e\text{-ph}}} = \frac{2\pi}{\hbar} \sum_{m,p} \int \frac{d\mathbf{q}}{\Omega_{\text{BZ}}} |\mathbf{M}_{n\mathbf{k},p\mathbf{q}}^{m\mathbf{k}+\mathbf{q}}|^2 \left[ \begin{array}{l} \left(f_{m\mathbf{k}+\mathbf{q}} + n_{p\mathbf{q}}\right) \delta(\varepsilon_{n\mathbf{k}} - \varepsilon_{m\mathbf{k}+\mathbf{q}} + \hbar\omega_{p\mathbf{q}}) \\ + \left(1 - f_{m\mathbf{k}+\mathbf{q}} + n_{p\mathbf{q}}\right) \delta(\varepsilon_{n\mathbf{k}} - \varepsilon_{m\mathbf{k}+\mathbf{q}} - \hbar\omega_{p\mathbf{q}}) \end{array} \right]. \quad (1)$$

Here, $\Omega_{\text{BZ}}$ is the volume of the first Brillouin zone. $\varepsilon_{n\mathbf{k}}$ represents the electron energy in band *n* at wavevector **k**, and $\omega_{p\mathbf{q}}$ represents the phonon frequency in mode *p* at wavevector **q**. $f_{n\mathbf{k}}$ and $n_{p\mathbf{q}}$ are the Fermi-Dirac and Bose-Einstein distribution functions, respectively. $\mathbf{M}_{n\mathbf{k},p\mathbf{q}}^{m\mathbf{k}+\mathbf{q}}$ is the electron-phonon coupling matrix indicating the coupling strength of the three-particle process *n***k** + *p***q** = *m***k** + **q**. The first term on the right-hand side of Eq. (1) is the scattering rate of electrons due to the absorption of a phonon, while the second term results from the phonon-emission process. Once the scattering rates are obtained, the mobility tensor can be calculated by solving the BTE, which is given by [8]

$$\mu_{\alpha\beta} = \frac{g_e e}{\Omega N_{\mathbf{k}} n_c} \sum_{n\mathbf{k}} v_{n\mathbf{k},\alpha} v_{n\mathbf{k},\beta} \tau_{n\mathbf{k}}^{e\text{-ph}} \frac{\partial f_{n\mathbf{k}}}{\partial \varepsilon_{n\mathbf{k}}}, \quad (2)$$

where $\Omega$, $N_{\mathbf{k}}$ and $g_e$ are, respectively, the volume of the unit cell, the number of sampling points in the Brillouin zone and the degeneracy of electrons. $n_c$ is the carrier concentration, and $v_{n\mathbf{k},\alpha}$ denotes the group velocity of electrons in the $\alpha$ direction.

The electron and phonon band structures are, respectively, computed within the density-functional theory (DFT) and DFPT framework, employing the Quantum ESPRESSO package [43]. We use a fully relativistic norm-conserving pseudopotential with the Perdew-Zunger exchange-correlation functional for all types of atoms in this study. In the DFT calculations, a 12×12×12 Monkhorst-Pack **k**-mesh associated with a plane-wave cutoff energy of 80 Ry is used for self-consistent and non-self-consistent field calculations. The optimized lattice constants of BP, BAs and BSb are 4.488, 4.763 and 5.179 Å, respectively, which are in good agreement with reported values [44]. The electron band structures are displayed in Fig. (1). In the DFPT calculations, the dynamical matrices are computed on a 6×6×6 **q**-mesh. It is worth noting that the convergence threshold is chosen as $10^{-20}$ Ry in DFPT calculations to guarantee well-converged phonon perturbed potentials. After the electron energies are determined, phonon frequencies and perturbed potentials are prepared. The EPW package [45] is employed to interpolate these quantities to 200×200×200 **k**-mesh and 100×100×100 **q**-mesh, thus obtaining the fine-meshed electron-phonon coupling matrices. For better convergence, we have added the tetrahedral method [46] into the original EPW code to treat the Brillouin-zone integration of Eq. (1). Finally, we calculate carrier mobility and MFP spectrum, enabled through our in-house code—a linearized BTE solver.

Figures 1(d) to 1(f) show the scattering rates of holes (left) and electrons (right panel) broken down into each phonon mode for BP, BAs and BSb at 300 K and a carrier concentration of $10^{17}$ cm$^{-3}$. It can be seen that among these materials, the electron-phonon interactions in both valence and conduction bands are dominated by the acoustic-deformation-potential scattering (particularly by the LA phonons). As expected of deformation-potential scattering, the electron scattering rate due to the acoustic mode basically follows the functional form of $\varepsilon^x$, where the exponent $x$ is determined by the band shape ($x = 0.5$ for parabolic dispersion) [8]. Through inspection of the plots of scattering rates, inter-valley versus intra-valley transitions can be identified. For BSb, the LA-phonon scattering rate shows a kink around 0.03 eV in the conduction band, which is due to inter-valley transitions. This occurs in BSb due to the shallow pocket of the band minimum (see Fig. 1(c)), where the intra-valley scatterings can only take place over a very small range of energy. In contrast, intra-valley transitions dominate the electron-phonon interaction in BP and BAs.

The dark red points (scattering due to LO phonons) in Figs. 1(d) to 1(f) reflect the electron interaction with the macroscopic polarization field in polar materials. As an example, in the conduction band of BAs (right panel in Fig. 1(e)), the polar-optical-phonon interaction is dominated by the phonon-absorption processes around the band edge, showing a nearly energy-independent function of the scattering rate. As the energy increases beyond $\hbar\omega_\text{LO}$ (the energy of LO phonons), the phonon-emission processes start to take over the scattering channel, and the polar scattering has substantial contribution to the electron-phonon interaction. However, in contrast to what occurs in strongly polar materials such as GaAs, the polar-optical-phonon interaction has much less effect on the electron transport in BAs (also similarly in BP and BSb), for which there are two possible reasons. (i) The polarization field induced by the LO phonons is weak due to either the large dielectric constant or the small effective charge, both of which tend to decrease the coupling strength between electrons and LO phonons. (ii) The occupation number of LO phonon is low, which usually occurs when the optical mode possesses high vibrational frequencies. We found that the suppression of polar scattering is caused by the latter, which will be discussed shortly.

The mobility can be obtained through Eq. (2) with the computed scattering rates. It should be emphasized that ionized impurity scattering is not included here. We found that including the spin-orbit coupling in the DFT calculation is crucial for studying hole mobility, since the presence of spin-splitting gap in the valence band manifold substantially affect the scattering channels of electron-phonon interactions. This is observed in the trend of total scattering rates as shown in Figs. 1(d) to 1(f)—the scattering rate of holes becomes smaller due to the larger spin-splitting gap as the anionic atom goes down the periodic table. In Fig. 2(a), the electron and hole mobilities of BP, BAs and BSb are presented at 300 K with respect to varying carrier concentrations. The mobilities drop at high carrier concentration since the higher-energy electrons are usually subjected to stronger phonon scatterings arising from larger joint density of states. The temperature-dependent mobilities are shown in Fig. 2(b), which indicates that the carriers' mobilities are sensitive to temperature. In our calculations, BAs shows a 85% reduction in hole mobility (from 6360 to 930 cm$^2$/V-s) when the temperature rises from 200 K to 400 K. This can be attributed to the increase in the phonon occupation number at elevated temperatures, which leads to enhanced electron-phonon interaction. In BAs, the temperature dependencies of hole and electron mobilities follow the trend of $\sim T^{-2.8}$ and $\sim T^{-2.4}$, respectively, which departs from the typical $T^{-1.5}$ trend. This is owed to the enhancement of inter-valley and inter-band scatterings at room temperatures [8].

Our first-principles calculations reveal that single-crystal BAs simultaneously possesses a high hole mobility (2110 cm$^2$/V-s) and high electron mobility (and 1400 cm$^2$/V-s) at room temperature. This is unusual since most common IV and III-V group semiconductors exhibit a high mobility for only one type of carrier. For example, InSb and GaAs are outstanding *n*-type conductors but are poor *p*-type conductors (77000 cm$^2$/V-s and 850 cm$^2$/V-s for InSb, and 9000 cm$^2$/V-s and 400 cm$^2$/V-s for GaAs at room temperature [44]). High *n*- and *p*-type mobilities obtained from one material is beneficial for most electronic devices, ranging from diodes to transistors and photovoltaic devices [1]. This characteristic, combined with outstanding thermal conductivity, long hot-carrier lifetime and suitable energy bandgap, makes BAs attractive for a wide range of electronic applications.

We further discuss the MFP spectrum of the electrical conductivity, which provides critical information for manipulating electron transport at the nanoscale. In Fig. 3(a), we plot the MFP spectra of holes (dashed) and electrons (solid lines) for BP, BAs and BSb at 300 K and a carrier concentration of $10^{17}$ cm$^{-3}$. Compared to the phonon MFP spectrum, the electron MFP spectrum is expected to have a much narrower span of distribution since the Fermi-Dirac distributions limit the states that can contribute to transport. For BAs, our calculations demonstrate that contributions to the hole conductivity mostly come from states with MFPs within the range of 50 nm to 180 nm, while for the electrical conductivity, they come from states with MFPs of 20 nm to 70 nm. In comparison, contributions to the thermal conductivity come from phonons with MFPs up to several micrometers [47]. In BAs, the spin-orbit-coupling-induced splitting leads to the isolation of the spin-off band (see Fig. 1(b), $E_{SO}$ = 0.216 eV by DFT calculation) from the valence band maximum, and therefore the hole transport mostly happens in the light- and heavy-hole bands. The spin-off band starts to participate in the electron-phonon interactions in heavily-doped regimes where the Fermi level shifts into the valence band. This drives the more significant shift (reduction) of the hole MFP spectrum compared to the electron MFP spectrum as the carrier concentration increases (Fig. 3(b)), and is the underlying reason behind the greater electron mobility in BAs than hole mobility, at high carrier concentration (Fig. 2(a)).

Finally, we discuss the physical mechanisms underlying the remarkable suppression of the polar-optical-phonon interaction in BAs. The coupling strength between phonons along the Γ→X and Γ→L directions, and between the electrons and holes at the band extrema, are plotted in Fig. 4(a). The coupling strength due to LO phonons diverge near the Γ point, which is expected for polar materials, as discovered by Fröhlich [5]. By DFPT calculations, we found that BAs has a comparable high-frequency dielectric constant ($\varepsilon_\infty$ = 9.6) to GaAs ($\varepsilon_\infty$ = 10.9). Although the calculated Born effective charges ($Z^*$) of BAs are ±0.55, about four times smaller than those of GaAs (note that the long-range Coulomb contribution in the Fröhlich interaction is proportional to $Z^*/\varepsilon_\infty$ [18]), the electron-LO-phonon coupling strength still overwhelms other modes, especially when **q** is small. For comparison, we plot the coupling strength of GaAs in Fig. 4(b). The results as shown indicate that the interactions between the carriers and the LO phonons should be significant; however, the electron-phonon interaction in BAs is dominated by acoustic phonons as previously discussed. This discrepancy leads to the investigation in the occupations of LO phonons.

In Fig. 4(c), we plot the room-temperature phonon occupancy function of BAs along the corresponding directions. One can see that the occupation number of acoustic phonons is at least two orders of magnitude greater than that of LO phonons around the $\Gamma$ point, which is clearly different from what can be seen for GaAs (Fig. 4(d)). This can be attributed to the high frequency of optical modes—the reduction of LO-phonon occupation directly suppresses the phase space of polar interaction, which leads to the lower electron-LO-phonon scattering rates. Another effect of the large LO-phonon frequency is that in polar materials, the electrons require a higher energy to assist the phonon-emission transitions, which means that the electrons near the band edge (for BAs, $\varepsilon < 0.09$ eV) are only subjected to the phonon-absorption scatterings as shown in Figs. 1(d) to 1(f). The presence of its high frequency also reduces the phonon occupation of optical modes at the Brillouin zone boundary, and therefore suppresses the inter-valley and inter-band transitions that are often mediated by optical phonons with large **q**. These findings offer a new insight for searching high mobility polar semiconductors by means of phonon band structure, in addition to the traditional perspectives such as electron group velocity, effective mass and deformation potential.

To summarize, we performed first-principles calculations to gain a deeper understanding of charge transport in the three boron-V compounds (BP, BAs and BSb). Single-crystal BAs is found to possess both high hole and electron mobilities, which can be attributed to its unique feature of lattice dynamics. The outstanding electrical performance as predicted by first-principles calculation stems from the existence of high optical phonon frequency, which leads to a large suppression of polar scattering as well as of inter-valley and inter-band transitions mediated by large-**q** optical phonons, due to their lower occupancy at room temperature. In addition to its high thermal conductivity, the simultaneously high hole and electron mobility presents itself as a new advantage for BAs as a material of great interest in electronic and photovoltaic applications.

This work was funded by the Multidisciplinary University Research Initiative (MURI) program, Office of Naval Research under a Grant No. N00014-16-1-2436 through the University of Texas at Austin.

**Figure captions**

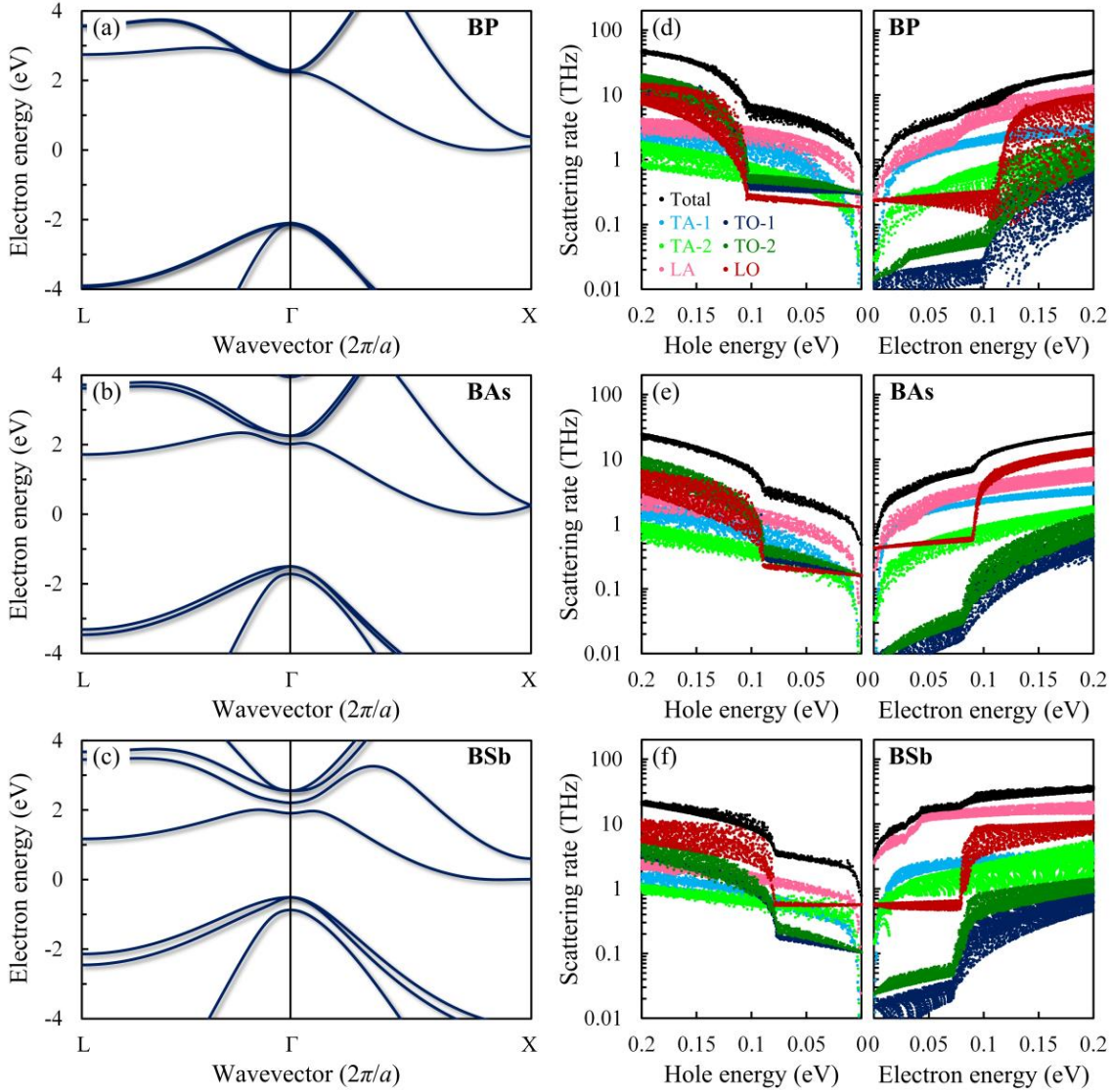

Figure 1. Electron band structures of (a) BP, (b) BAs and (c) BSb obtained by DFT calculations. The bandgaps have been expanded to the experimental values (BP = 2.10 eV, BAs = 1.50 eV and BSb = 0.51 eV [44]) according to the rigid band approximation in order to determine more reasonable Fermi levels for the calculations of transport properties. Carrier scattering rates decomposed into different phonon branches for (d) BP, (e) BAs and (f) BSb within 0.2 eV from the band extrema. The left and right panels represent the scattering rates of holes and electrons, respectively.

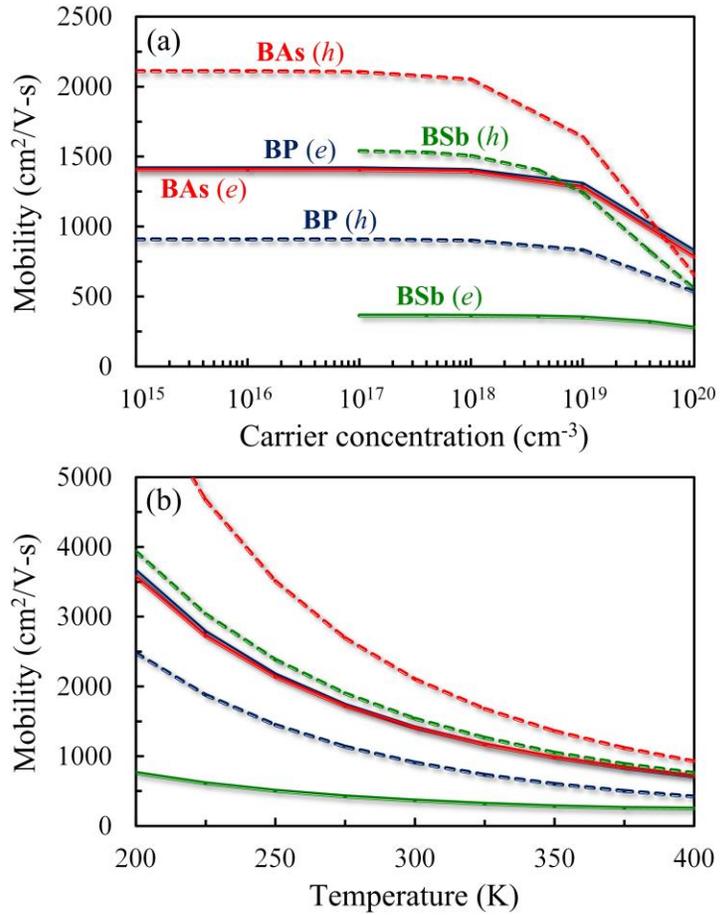

Figure 2. Carrier mobilities of BP (blue), BAs (red) and BSb (green) with varying (a) carrier concentrations and (b) temperatures. The dashed and solid lines represent the mobilities of holes and electrons, respectively. The calculated intrinsic carrier concentration for BP, BAs and BSb are 69.3, $4.7\times10^6$, and $1.1\times10^{15}$ cm$^{-3}$, respectively.

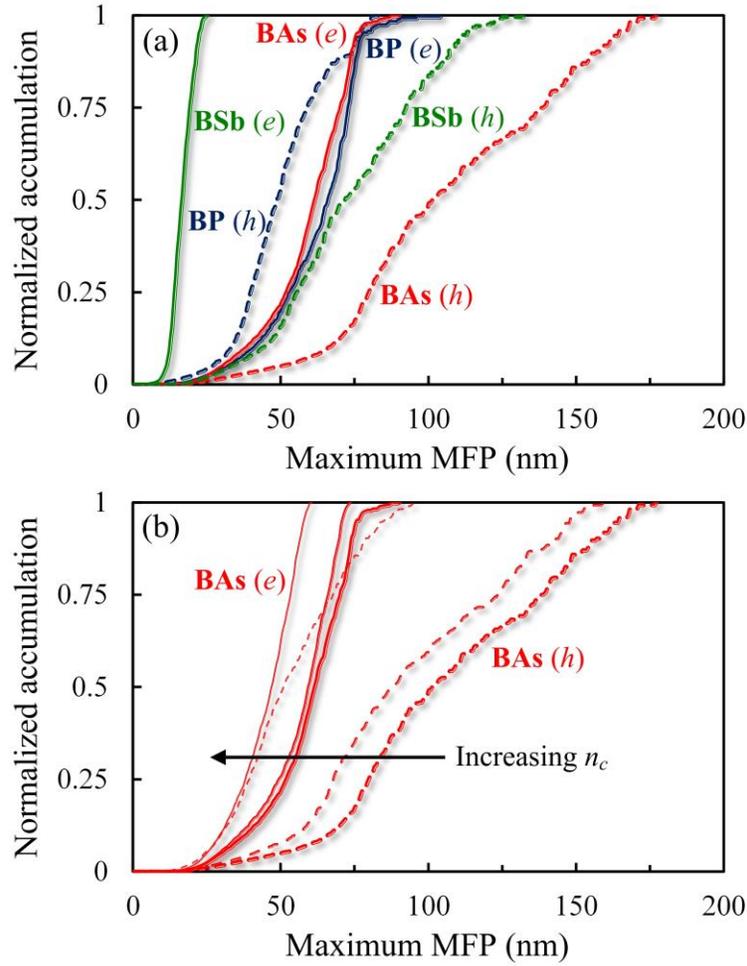

Figure 3. (a) Accumulations of electrical conductivity of BP (blue), BAs (red) and BSb (green) at 300 K and a carrier concentration of $10^{17}$ cm$^{-3}$, normalized by the corresponding maximum MFPs. The dashed and solid lines represent the MFP spectrum of holes and electrons, respectively. (b) Accumulations of electrical conductivity of BAs at different carrier concentrations. The thick to thin lines represent carrier concentrations of $10^{17}$, $10^{19}$ and $10^{20}$ cm$^{-3}$, respectively.

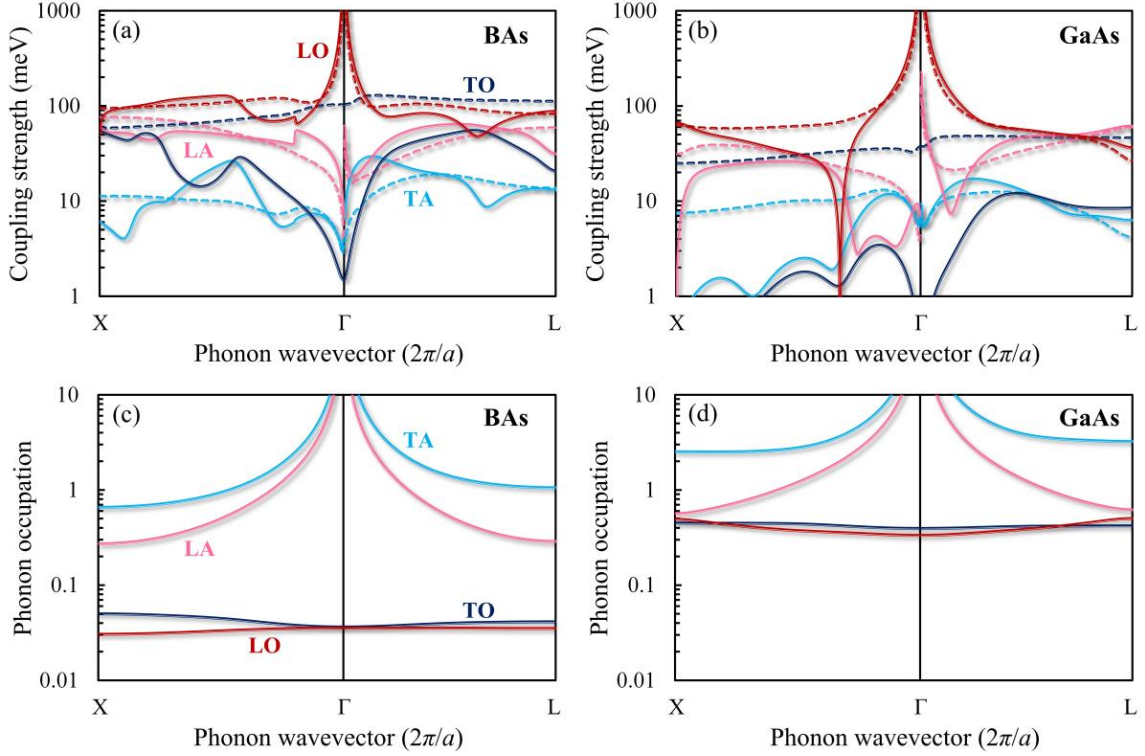

Figure 4. Coupling strengths between phonons along two high-symmetry directions (Γ→X and Γ→L) and holes (dashed) and electrons (solid lines) at band extrema for (a) BAs and (b) GaAs. The transverse-acoustic (TA) and TO modes are degenerate along the two high-symmetry directions, respectively. The coupling strength of LA phonons in Γ→L direction diverges as **q**→0, arising from piezoelectric interactions since the inversion symmetry is broken in both materials [8]. The corresponding phonon occupations, $n_{p\mathbf{q}} = [\exp(\hbar\omega_{p\mathbf{q}}/k_B T) - 1]^{-1}$, of BAs and GaAs are plotted in (c) and (d), respectively.